\documentclass[10pt]{article}

\usepackage{cite}

\begin{document}

\vspace{2cm}
Preprint hep-ph/0006089

\vspace{3cm}

\begin{center}
{\Large \sf Improved Conformal Mapping of the Borel Plane}
\end{center}

\vspace{1cm}

\begin{center}
U.~D.~Jentschura and G.~Soff
\end{center}

\vspace{0.2cm}

\begin{center}
{\it Institut f\"ur Theoretische Physik, TU Dresden, 01062
Dresden, Germany}\\[2ex]
{\bf Email:} jentschura@physik.tu-dresden.de, soff@physik.tu-dresden.de
\end{center}

\vspace{1.3cm}

\begin{center}
\begin{minipage}{10.5cm}
{\underline{Abstract}}
The conformal mapping of the Borel plane can be
utilized for the analytic continuation of the Borel transform to the
entire positive real semi-axis and is thus helpful in the resummation 
of divergent perturbation series in quantum field theory. We observe that
the convergence can be accelerated
by the application of Pad\'{e} approximants to the Borel transform
expressed as a function of the conformal variable, i.e.~by a
combination of the analytic continuation via conformal mapping and a
subsequent numerical approximation by rational approximants. 
The method is primarily useful in those cases 
where the leading (but not sub-leading) 
large-order asymptotics of the perturbative coefficients are known.
\end{minipage}
\end{center}

\vspace{1.3cm}

Asymptotic problems and properties
\newpage

The problem of the resummation of quantum field theoretic series is of
obvious importance in view of the divergent, asymptotic character of
the perturbative expansions~\cite{LGZJ1990,ZJ1996,Fi1997}.  The
convergence can be accelerated when additional information is
available about large-order asymptotics of the perturbative
coefficients~\cite{JeWeSo2000}. In the example cases discussed
in~\cite{JeWeSo2000}, the location of several poles in the Borel
plane, known from the leading and next-to-leading large-order
asymptotics of the perturbative coefficients, is utilized in order to
construct specialized resummation prescriptions.  Here, we consider a
particular perturbation series, investigated in~\cite{BrKr1999}, where
only the {\em leading} large-order asymptotics of the perturbative
coefficients are known to sufficient accuracy, and the subleading
asymptotics have -- not yet -- been determined. Therefore, the
location of only a single pole -- the one closest to the origin -- in
the Borel plane is available. In this case, as discussed
in~\cite{CaFi1999,CaFi2000}, the (asymptotically optimal) conformal
mapping of the Borel plane is an attractive method for the analytic
continuation of the Borel transform beyond its circle of convergence
and, to a certain extent, for accelerating the convergence of the
Borel transforms.  Here, we argue that the convergence of the
transformation can be accelerated further
when the Borel transforms, expressed as
a function of the conformal variable which mediates the analytic
continuation, are additionally convergence-accelerated by the
application of Pad\'{e} approximants.

First we discuss, in general terms, the construction of the improved
conformal mapping of the Borel plane which is used for the
resummation of the perturbation series defined in 
Eqs.~(\ref{gammaPhi4}) and~(\ref{gammaYukawa}) 
below. The method uses as input data the numerical 
values of a finite number of perturbative coefficients and
the leading large-order asymptotics of the perturbative coefficients,
which can, under appropriate circumstances,
be derived from an empirical investigation of a finite number of 
coefficients, as it has been done in~\cite{BrKr1999}. 
We start from an asymptotic, divergent perturbative expansion
of a physical observable $f(g)$ in powers of a coupling parameter $g$,
\begin{equation}
\label{power}
f(g) \sim \sum_{n=0}^{\infty} c_n\,g^n\,,
\end{equation}
and we consider the generalized Borel transform of the 
$(1,\lambda)$-type (see Eq.~(4) in~\cite{JeWeSo2000}),
\begin{equation}
\label{BorelTrans}
f^{(\lambda)}_{\rm B}(u) \; \equiv \;
f^{(1,\lambda)}_{\rm B}(u) \; = \; 
\sum_{n=0}^{\infty} \frac{c_n}{\Gamma(n+\lambda)}\,u^n\,.
\end{equation}
The full physical solution can be reconstructed
from the divergent series~(\ref{power}) by evaluating the
Laplace-Borel integral, which is defined as
\begin{equation}
\label{BorelIntegral}
f(g) = \frac{1}{g^\lambda} \,
\int_0^\infty {\rm d}u \,u^{\lambda - 1} \, 
\exp\bigl(-u/g\bigr)\,
f^{(\lambda)}_{\rm B}(u)\,.
\end{equation}
The integration variable $u$ is referred to as the Borel variable.
The integration is carried out either along the real axis or
infinitesimally above or below it (if Pad\'{e} approximants are used for
the analytic continuation, modified
integration contours have been proposed~\cite{Je2000}). The most
prominent issue in the theory of the Borel resummation is the
construction of an analytic continuation for the 
Borel transform~(\ref{BorelTrans}) from a
finite-order partial sum of the
perturbation series (\ref{power}), which we denote by
\begin{equation}
\label{PartialSum}
f^{(\lambda),m}_{\rm B}(u) =
\sum_{n=0}^{m} \frac{c_n}{\Gamma(n+\lambda)}\,u^n\,.
\end{equation}
The analytic continuation can be accomplished using the direct
application of Pad\'{e} approximants to the partial sums 
of the Borel transform 
$f^{(\lambda),m}_{\rm B}(u)$~\cite{BrKr1999,Je2000,Raczka1991,Pi1999} or by
a conformal mapping~\cite{SeZJ1979,LGZJ1983,GuKoSu1995,CaFi1999,CaFi2000}. 
We now assume that the {\em leading} large-order asymptotics of 
the perturbative
coefficients $c_n$ defined in Eq.~(\ref{power}) is factorial, 
and that the coefficients display an alternating sign pattern. 
This indicates the existence of a singularity
(branch point) along the negative real axis
corresponding to the leading large-order growth of the 
perturbative coefficients, which we assume to be at $u=-1$.
For Borel transforms which have only a single cut in the complex plane 
which extends from $u=-1$ to $u=-\infty$, the following
conformal mapping has been recommended as optimal~\cite{CaFi1999},
\begin{equation}
\label{DefZ}
z = z(u) = \frac{\sqrt{1+u}-1}{\sqrt{1+u}+1}\,.
\end{equation}
Here, $z$ is referred to
as the conformal variable. The cut Borel plane is mapped
unto the unit circle by the conformal mapping (\ref{DefZ}).
We briefly mention that a large variety of similar
conformal mappings have been discussed in
the literature~\cite{BrBaIlFlTaSm1994,AlNaRi1995,SoSu1996,
ChKuSt1996a,ChKuSt1996b,ChHaSt1998}. 

It is worth noting that 
conformal mappings which are adopted for doubly-cut Borel
planes have been discussed in~\cite{CaFi1999,CaFi2000}.
We do not claim here that 
it would be impossible to construct conformal mappings which reflect
the position of more than two renormalon poles or branch points
in the complex plane.
However, we stress that such a conformal mapping is likely to have a 
more complicated mathematical structure than, for example, 
the mapping defined in Eq.~(27) in~\cite{CaFi1999}. Using the alternative
methods described in~\cite{JeWeSo2000}, poles (branch points)
in the Borel plane
corresponding to the subleading asymptotics can be incorporated easily
provided their position in the Borel plane is known. In a concrete
example (see Table~1 in~\cite{JeWeSo2000}), 14 poles in the Borel plane have
been fixed in the denominator of the Pad\'{e} approximant constructed
according to Eqs.~(53)--(55) in~\cite{JeWeSo2000}, and accelerated 
convergence of the transforms is observed. 
In contrast to the investigation~\cite{JeWeSo2000}, we assume here that only 
the {\em leading} large-order factorial asymptotics of the perturbative
coefficients are known.

We continue with the discussion of the conformal mapping (\ref{DefZ}).
It should be noted that for series whose leading singularity 
in the Borel plane is at
$u = -u_0$ with $u_0 > 0$, an appropriate rescaling of the Borel
variable $u \to |u_0|\, u$ is necessary on the right-hand side
of Eq.~(\ref{BorelIntegral}).
Then, $f^{(\lambda)}_{\rm B}(|u_0|\,u)$ as a function of $u$ has its
leading singularity at $u = -1$ (see also Eq.~(41.57) in~\cite{ZJ1996}).
The Borel integration
variable $u$ can be expressed as a function of $z$ as follows,
\begin{equation}
\label{UasFuncOfZ}
u(z) = \frac{4 \, z}{(z-1)^2}\,.
\end{equation}
The $m$th partial sum of the Borel transform (\ref{PartialSum}) can be
rewritten, upon expansion of the $u$ in powers of $z$, as
\begin{equation}
\label{PartialSumConformal}
f^{(\lambda),m}_{\rm B}(u) = 
f^{(\lambda),m}_{\rm B}\bigl(u(z)\bigr) = 
\sum_{n=0}^{m} C_n\,z^n + {\cal O}(z^{m+1})\,,
\end{equation}
where the coefficients $C_n$ as a function of the $c_n$ are uniquely
determined (see, e.g., Eqs.~(36) and (37) of~\cite{CaFi1999}). We
define partial sum of the Borel transform, expressed as a function of
the conformal variable $z$, as
\begin{equation}
f'^{(\lambda),m}_{\rm B}(z) = \sum_{n=0}^{m} C_n\,z^n\,.
\end{equation}
In a previous investigation~\cite{CaFi1999}, Caprini and Fischer
evaluate the following transforms,
\begin{equation}
\label{CaFiTrans}
{\cal T}'_m f(g) = \frac{1}{g^\lambda}\,
\int_0^\infty {\rm d}u \,u^{\lambda - 1} \,\exp\bigl(-u/g\bigr)\,
f'^{(\lambda),m}_{\rm B}(z(u))\,.
\end{equation}
Caprini and Fischer~\cite{CaFi1999} observe the apparent numerical
convergence with increasing $m$. The limit as $m\to\infty$, provided
it exists, is then assumed to represent the complete, physically
relevant solution,
\begin{equation}
f(g) = \lim_{m\to\infty} {\cal T}'_m f(g)\,.
\end{equation}
We do not consider the question of the existence of this limit here
(for an outline of questions related to these issues we refer
to~\cite{CaFi2000}).  

In the absence of further information on the analyticity domain
of the Borel transform~(\ref{BorelTrans}), we cannot necessarily
conclude that $f^{(\lambda)}_{\rm B}{\mathbf (}u(z){\mathbf )}$
as a function of $z$ is analytic inside the unit circle of the 
complex $z$-plane, or that, for example, the conditions of 
Theorem~5.2.1 of~\cite{BaGr1996} are fulfilled. Therefore,
we propose a modification of the
transforms (\ref{CaFiTrans}). In particular, we advocate the
evaluation of (lower-diagonal) Pad\'{e} 
approximants~\cite{BaGr1996,BeOr1978} to the function
$f'^{(\lambda),m}_{\rm B}(z)$, expressed as a function of $z$,
\begin{equation}
\label{ConformalPade}
f''^{(\lambda),m}_{\rm B}(z) = 
\bigg[ [\mkern - 2.5 mu [m/2] \mkern - 2.5 mu ] \bigg/
[\mkern - 2.5 mu [(m+1)/2] \mkern - 2.5 mu ]
\bigg]_{f'^{(\lambda),m}_{\rm B}}\!\!\!\left(z\right)\,.
\end{equation}
We define the following transforms,
\begin{equation}
\label{AccelTrans}
{\cal T}''_m f(g) = \frac{1}{g^\lambda}\,
\int_{C_j} {\rm d}u \,u^{\lambda - 1} \,\exp\bigl(-u/g\bigr)\,
f''^{(\lambda),m}_{\rm B}\bigl(z(u)\bigr)
\end{equation}
where the integration contour $C_j$ ($j=-1,0,1$) have been
defined in~\cite{Je2000}. These integration contours
have been shown to to provide
the physically correct analytic continuation of resummed
perturbation series for those cases where the evaluation of the
standard Laplace-Borel integral~(\ref{BorelIntegral}) is impossible
due to an insufficient analyticity domain of the integrand
(possibly due to multiple branch cuts) or due to
spurious singularities in view of the finite order of the 
Pad\'{e} approximations defined in~(\ref{ConformalPade}). We should mention 
potential complications due to multi-instanton contributions,
as discussed for example in Ch.~43 of~\cite{ZJ1996} 
(these are not encountered in the current investigation).
In this letter, we use exclusively the contour 
$C_0$ which is defined as the half sum of the contours
$C_{-1}$ and $C_{+1}$ displayed in Fig.~1 in~\cite{Je2000}.
At increasing $m$, the limit as $m\to\infty$, provided it exists, is
then again assumed to represent the complete, physically relevant
solution,
\begin{equation}
f(g) = \lim_{m\to\infty} {\cal T}''_m f(g)\,.
\end{equation}
Because we take advantage of the special integration contours $C_j$, 
analyticity of the Borel transform
$f^{(\lambda)}_{\rm B}{\mathbf (}u(z){\mathbf )}$
inside the unit circle of the complex $z$-plane is not required,
and additional acceleration of the convergence is mediated by employing
Pad\'{e} approximants in the conformal variable $z$.

%
%
\begin{table}[tbh]
\begin{center}
\begin{minipage}{15cm}
\begin{center}
\caption{\label{table1} Resummation of the perturbation series
(\ref{gammaPhi4}) for the anomalous dimension $\gamma$ function of the
6-dimensional $\phi^3$ theory by the method defined in
Eqs.~(\ref{power})--(\ref{AccelTrans}).  The transforms ${\cal T}''_m
\gamma_{\rm hopf}(g)$ are shown in the transformation order
$m=28,~29,~30$. The coupling $g$ assumes the values $g=5.0,~5.5,~6.0$
and $g=10.0$.}
\vspace*{0.3cm}
\begin{tabular}{cr@{.}lr@{.}lr@{.}lr@{.}l}
\hline
\hline
\rule[-3mm]{0mm}{8mm} $m$ &
\multicolumn{2}{c}{${\cal T}''_m \gamma_{\rm hopf}(5.0)$} &
\multicolumn{2}{c}{${\cal T}''_m \gamma_{\rm hopf}(5.5)$} &
\multicolumn{2}{c}{${\cal T}''_m \gamma_{\rm hopf}(6.0)$} &
\multicolumn{2}{c}{${\cal T}''_m \gamma_{\rm hopf}(10.0)$} \\
\hline
\rule[-3mm]{0mm}{8mm}
28 &
 $-0$ & $501~565~232$ &
 $-0$ & $538~352~234$ &
 $-0$ & $573~969~740$ &
 $-0$ & $827~506~173$ \\
\rule[-3mm]{0mm}{8mm}
 29 &
 $-0$ & $501~565~232$ &
 $-0$ & $538~352~233$ &
 $-0$ & $573~969~738$ &
 $-0$ & $827~506~143$ \\
\rule[-3mm]{0mm}{8mm}
30 &
 $-0$ & $501~565~231$ &
 $-0$ & $538~352~233$ &
 $-0$ & $573~969~738$ &
 $-0$ & $827~506~136$ \\
\hline
\hline
\end{tabular}
\end{center}
\end{minipage}
\end{center}
\end{table}

%
%
\begin{table}[tbh]
\begin{center}
\begin{minipage}{15cm}
\begin{center}
\caption{\label{table2} Resummation of the perturbation series
(\ref{gammaYukawa}) for the anomalous dimension of the Yukawa coupling
via the method defined in Eqs.~(\ref{power})--(\ref{AccelTrans}).  The
transforms ${\cal T}''_m {\tilde \gamma}_{\rm hopf}(g)$ are shown in
the order of transformation $m=28,~29,~30$. For the coupling $g$, we
consider the values $g=5.0,~5.5,~6.0$ and $g=30^2/(4 \pi)^2 = 5.69932\dots$.}
\vspace*{0.3cm}
\begin{tabular}{cr@{.}lr@{.}lr@{.}lr@{.}l}
\hline
\hline
\rule[-3mm]{0mm}{8mm} $m$ &
\multicolumn{2}{c}{${\cal T}''_m {\tilde \gamma}_{\rm hopf}(5.0)$} &
\multicolumn{2}{c}{${\cal T}''_m {\tilde \gamma}_{\rm hopf}(5.5)$} &
\multicolumn{2}{c}{${\cal T}''_m {\tilde \gamma}_{\rm hopf}(6.0)$} &
\multicolumn{2}{c}{${\cal T}''_m 
  {\tilde \gamma}_{\rm hopf}\bigl(30^2/(4\pi)^2\bigr)$} \\
\hline
\rule[-3mm]{0mm}{8mm}
28 &
 $-1$ & $669~071~213$ &
 $-1$ & $800~550~588$ &
 $-1$ & $928~740~624$ &
 $-1$ & $852~027~809$ \\
\rule[-3mm]{0mm}{8mm}
 29 &
 $-1$ & $669~071~214$ &
 $-1$ & $800~550~589$ &
 $-1$ & $928~740~626$ &
 $-1$ & $852~027~810$ \\
\rule[-3mm]{0mm}{8mm}
30 &
 $-1$ & $669~071~214$ &
 $-1$ & $800~550~589$ &
 $-1$ & $928~740~625$ &
 $-1$ & $852~027~810$ \\
\hline
\hline
\end{tabular}
\end{center}
\end{minipage}
\end{center}
\end{table}

We consider the resummation of two particular perturbation series
discussed in~\cite{BrKr1999} for the anomalous dimension $\gamma$
function of the $\phi^3$ theory in 6 dimensions
and the Yukawa coupling in 4
dimensions. The perturbation series for the $\phi^3$ theory is given
in Eq.~(16) in~\cite{BrKr1999},
\begin{equation}
\label{gammaPhi4}
\gamma_{\rm hopf}(g) \sim 
  \sum_{n=1}^{\infty} (-1)^n \, \frac{G_n}{6^{2 n - 1}} \, g^n\,,
\end{equation}
where the coefficients
$G_n$ are given in Table~1 in~\cite{BrKr1999} for
$n=1,\dots,30$ (the $G_n$ are real and positive). 
We denote the coupling parameter $a$ used
in~\cite{BrKr1999} as $g$; this is done in order to ensure
compatibility with the general power series given in
Eq.~(\ref{power}). Empirically, Broadhurst and Kreimer derive the
large-order asymptotics
\begin{equation}
G_n \sim {\rm const.} \; \times \;
12^{n-1} \, \Gamma(n+2)\,, \qquad n\to\infty\,,
\end{equation}
by investigating the explicit numerical values of the coefficients
$G_1,\dots,G_{30}$. The leading asymptotics of the perturbative
coefficients $c_n$ are therefore (up to a constant prefactor)
\begin{equation}
\label{LeadingPhi4}
c_n \sim (-1)^n \frac{\Gamma(n+2)}{3^n}\,, \qquad n\to\infty\,.
\end{equation}
This implies that the $\lambda$-parameter in the Borel transform
(\ref{BorelTrans}) should be set to $\lambda=2$ (see also the notion
of an asymptotically optimized Borel transform discussed
in~\cite{JeWeSo2000}). In view of Eq.~(\ref{LeadingPhi4}), the pole
closest to the origin of the Borel transform (\ref{BorelTrans}) is
expected at
\begin{equation}
u = u^{\rm hopf}_0 = -3\,,
\end{equation}
and a rescaling of the Borel variable $u \to 3\,u$ in
Eq.~(\ref{BorelIntegral}) then leads to an expression to which the
method defined in Eqs.~(\ref{power})--(\ref{AccelTrans}) can be applied
directly. For the Yukawa coupling, the $\gamma$-function reads
\begin{equation}
\label{gammaYukawa}
{\tilde \gamma}_{\rm hopf}(g) \sim
  \sum_{n=1}^{\infty} (-1)^n \, 
     \frac{{\tilde G}_n}{2^{2 n - 1}} \, g^n\,,
\end{equation}
where the ${\tilde G}_n$ are given in Table~2 in~\cite{BrKr1999} for
$n=1,\dots,30$. Empirically, i.e.~from an investigation of the
numerical values of ${\tilde G}_1,\dots,{\tilde G}_{30}$, the following 
factorial growth in large order is derived~\cite{BrKr1999},
\begin{equation}
{\tilde G}_n \sim {\rm const.'} \; \times \;
2^{n-1} \, \Gamma(n+1/2)\,, \qquad n\to\infty\,.
\end{equation}
This leads to the following asymptotics for the perturbative 
coefficients (up to a constant prefactor),
\begin{equation}
c_n \sim (-1)^n \frac{\Gamma(n+1/2)}{2^n} \,, \qquad n\to\infty\,.
\end{equation}
This implies that an asymptotically optimal choice~\cite{JeWeSo2000} 
for the $\lambda$-parameter in (\ref{BorelTrans}) is $\lambda=1/2$.
The first pole of the Borel transform (\ref{BorelTrans}) is therefore
expected at
\begin{equation}
u = {\tilde u}^{\rm hopf}_0 = -2\,.
\end{equation}
A rescaling of the Borel variable according to $u \to 2\,u$ in
(\ref{BorelIntegral}) enables the application of the resummation
method defined in Eqs.~(\ref{power})--(\ref{AccelTrans}).

In Table~\ref{table1}, numerical values for the transforms ${\cal
T}''_m \gamma_{\rm hopf}(g)$ are given, which have been evaluated
according to Eq.~(\ref{AccelTrans}). The transformation order is in
the range $m=28~,29,~30$, and we consider coupling parameters
$g=5.0,~5.5,~6.0$ and $g=10.0$. The numerical values of the transforms display
apparent convergence to about 9 significant figures for $g \leq 6.0$
and to about 7 figures for $g=10.0$.  In
Table~\ref{table2}, numerical values for the transforms ${\cal T}''_m
{\tilde \gamma}_{\rm hopf}(g)$ calculated according to
Eq.~(\ref{AccelTrans}) are shown in the range $m=28,~29,~30$ for
(large) coupling strengths $g=5.0,~5.5,~6.0$. Additionally, the value
$g = 30^2/(4\,\pi)^2 = 5.69932\dots$ is considered as a special case
(as it has been done in~\cite{BrKr1999}).  Again, the numerical values
of the transforms display apparent convergence to about 9 significant
figures. At large coupling $g = 12.0$, the apparent convergence of the 
transforms suggests the following values: $\gamma_{\rm hopf}(12.0) =
-0.939\,114\,3(2)$ and ${\tilde \gamma}_{\rm hopf}(12.0) = 
-3.287\,176\,9(2)$.
The numerical results for the Yukawa case, i.e. for the function ${\tilde
\gamma}_{\rm hopf}$, have recently been confirmed by an improved analytic,
nonperturbative investigation~\cite{BrKr2000prep}  
which extends the perturbative
calculation~\cite{BrKr1999}.

We note that the transforms ${\cal T}'_m \gamma_{\rm hopf}(g)$ 
and ${\cal T}'_m {\tilde \gamma}_{\rm hopf}(g)$ 
calculated according to Eq.~(\ref{CaFiTrans}),
i.e.~by the unmodified conformal mapping, typically
exhibit apparent convergence to 5--6 significant figures in
the transformation order $m=28,~29,~30$ and at
large coupling $g \geq 5$. Specifically, the numerical values 
for $g=5.0$ are
\begin{eqnarray}
{\cal T}'_{28} \gamma_{\rm hopf}(g = 5.0) \; &=& \;
   -0.501~567~294\,, \nonumber\\[2ex]
{\cal T}'_{29} \gamma_{\rm hopf}(g = 5.0) \; &=& \;
   -0.501~564~509\,, \nonumber\\[2ex]
{\cal T}'_{30} \gamma_{\rm hopf}(g = 5.0) \; &=& \;
   -0.501~563~626\,. \nonumber
\end{eqnarray}
These results, when compared to the data in Table~\ref{table1},
exemplify the acceleration of the convergence by the additional
Pad\'{e} approximation of the Borel transform {\em expressed as a
function of the conformal variable} [see Eq.~(\ref{ConformalPade})].

It is not claimed here that the resummation method defined in
Eqs.~(\ref{power})--(\ref{AccelTrans}) necessarily provides the
fastest possible rate of convergence for the perturbation series
defined in Eq.~(\ref{gammaPhi4}) and (\ref{gammaYukawa}). Further
improvements should be feasible, especially if particular properties
of the input series are known and exploited (see in part the methods
described in~\cite{JeWeSo2000}). We also note possible improvements
based on a large-coupling expansion~\cite{We1996d}, in particular for
excessively large values of the coupling parameter $g$, or methods
based on order-dependent mappings (see~\cite{SeZJ1979,LGZJ1983}
or the discussion following Eq.~(41.67) in~\cite{ZJ1996}). 

The conformal mapping~\cite{CaFi1999,CaFi2000} is capable of
accomplishing the analytic continuation of the Borel transform
(\ref{BorelTrans}) beyond the circle of convergence. Pad\'{e}
approximants, applied directly to the partial sums of the Borel
transform~(\ref{PartialSum}), provide an alternative to this
method~\cite{Raczka1991,Pi1999,BrKr1999,Je2000,JeWeSo2000}. Improved
rates of convergence can be achieved when the convergence of the
transforms obtained by conformal mapping 
in Eq.~(\ref{PartialSumConformal})
is accelerated by evaluating Pad\'{e} approximants as in
Eq.~(\ref{ConformalPade}), and conditions on analyticity
domains can be relaxed in a favorable way when these methods are combined 
with the integration contours from Ref.~\cite{Je2000}. 
Numerical results for the resummed values of the 
perturbation series~(\ref{gammaPhi4}) and (\ref{gammaYukawa}) 
are provided in the
Tables~\ref{table1} and~\ref{table2}. By the improved conformal
mapping and other optimized resummation techniques (see,
e.g., the methods introduced in Ref.~\cite{JeWeSo2000}) 
the applicability of perturbative
(small-coupling) expansions can be generalized to the regime of large
coupling and still lead to results of relatively high accuracy.\\[4ex]

U.J.~acknowledges helpful conversations with E.~J.~Weniger,
I.~N\'{a}ndori, S.~Roether and P.~J.~Mohr. 
G.S.~acknowledges continued support from BMBF, DFG and GSI.

\end{document}